\documentclass{emulateapj}

\usepackage{natbib}

 \shorttitle{Oxygen abundance calibrations and N/O abundances of SDSS  galaxies}
 \shortauthors{Liang et al.}

\begin{document}

\title{The oxygen abundance calibrations and N/O abundance ratios of $\sim$
40,000 SDSS star-forming galaxies}
\author{Y. C. Liang\altaffilmark{1}, S. Y. Yin\altaffilmark{1,3}, F. Hammer
\altaffilmark{2}, L. C. Deng\altaffilmark{1}, H. Flores\altaffilmark{2}, and
B. Zhang\altaffilmark{1,3}}
\email{ycliang@bao.ac.cn}

\begin{abstract}
Using a large sample of 38,478 star-forming galaxies selected from the
Second Data Release of the Sloan Digital Sky Survey database (SDSS-DR2), we
derive analytical calibrations for oxygen abundances from several
metallicity-sensitive emission-line ratios: [N~\textsc{ii}]/H$\alpha$, 
[O~\textsc{iii}]/[N~\textsc{ii}], [N~\textsc{ii}]/[O~\textsc{ii}], 
[N~\textsc{ii}]/[S~\textsc{ii}], 
[S~\textsc{ii}]/H$\alpha$, and [O~\textsc{iii}]/H$\beta$. 
This consistent set of strong-line oxygen abundance calibrations will be
  useful for future abundance studies.
Among these calibrations, 
[N~\textsc{ii}]/[O~\textsc{ii}] is the best for metal-rich galaxies due to 
its independence on ionization parameter and low scatter.
Dust extinction must be
considered properly at first. These calibrations are more suitable for
metal-rich galaxies (8.4$<$12+log(O/H)$<$9.3), and for the nuclear regions of galaxies.
The observed relations are consistent with those expected from the
photoionization models of Kewley \& Dopita (2002). However, most
of the observational data spread in a range of ionization parameter $q$
from $1\times 10^7$ to $8\times 10^7$ cm s$^{-1}$, corresponding to log$U$=$-
$3.5 to $-$2.5, narrower than that suggested by the models. 
We also estimate the (N/O)
abundance ratios of this large sample of galaxies, and these are consistent
with the combination of a ``primary" and a dominant ``secondary" components
of nitrogen.
\end{abstract}


\altaffiltext{1}{National Astronomical Observatories, Chinese Academy of Sciences,
  20A Datun Road, Chaoyang District, Beijing 100012, P.R. China}  
\altaffiltext{2}{GEPI, Observatoire de Paris-Meudon, 92195 Meudon, France }  
\altaffiltext{3}{Department of Physics, Hebei Normal University,
Shijiazhuang 050016, P.R. China }



\section{Introduction}

The chemical properties of stars and gas within a galaxy are a fossil
record chronicling its history of star formation. 
Optical emission lines from H~\textsc{ii} regions
have long been the primary means of gas-phase chemical diagnostics in
galaxies (Aller 1942; Searle 1971; reviews by Peimbert 1975; Pagel 1986;
Shields 1990; Aller 1990; Peimbert, Rayo, \& Torres-Peimbert 1975;
Osterbrock 1989; de Robertis 1987). Osterbrock (1989) thoroughly
discusses the standard techniques for measuring the chemical properties of
ionized gas, which require measurements of H and He recombination lines
along with collisionally excited lines from one or more ionization states
of heavy element species.

Oxygen is the most commonly used metallicity indicator in the
interstellar medium (ISM) by virtue of its relatively high abundance and
strong emission lines in the optical region of the spectrum (e.g.,
[O~\textsc{ii}]$\lambda $ 3727 and [O~\textsc{iii}]$\lambda $$\lambda
$4959, 5007). Ideally, the oxygen abundance is measured directly from
ionic abundances obtained through a determination of the electron
temperature ($T_{e}$) of the  H~\textsc{ii} region. This is known as the
\textquotedblleft $T_{e}-$method"  (Pagel et al. 1992; Skillman \&
Kennicutt 1993). The ratio of a high-excitation auroral line such as
[O~\textsc{iii}]$\lambda $4363 to the lower excitation
[O~\textsc{iii}]$\lambda $$\lambda $4959,5007 lines provides a direct
measurement of the electron temperature in the physical medium where
O$^{++}$ is the dominant ionization state. 
However, [O~\textsc{iii}]$\lambda $4363 is only
prominent in metal-poor H~\textsc{ii} regions. In most cases, 
where [O~\textsc{iii}]$\lambda $4363
is too weak to be detected, metallicities of H~\textsc{ii} regions
have to be estimated from \textquotedblleft strong-line"
ratios. A commonly used one is $R_{23}$: 
\begin{equation}
R_{23}=\frac{f([\mathrm{OIII}]\lambda \lambda 4959,5007)+f([\mathrm{
OII]\lambda \lambda 3726,3729)}}{f(\mathrm{H\beta )}}. 
\end{equation}
Many calibrations of converting $R_{23}$ to 12+log(O/H) abundance have
been published, including Pagel et al. (1979), Pagel, Edmunds, \& Smith
(1980), Edmunds \& Pagel (1984), McCall, Rybski, \& Shields (1985),
Dopita \& Evans (1986), Torres-Peimbert, Peimbert, \& Fierro (1989),
Skillman, Kennicutt, \& Hodge (1989), McGaugh (1991), Zaritsky,
Kennicutt, \& Huchra (1994), Kobulnicky et al. (1999, K99), Pilyugin
(2000, 2001a,b), Tremonti et al. (2004, T04), Salzer et al. (2005) and
Kewley \& Dopita (2002) (hereafter KD02) etc. Some calibrations have
attempted to take the ionization parameters into account (e.g., McGaugh
1991; KD02 etc.), but others have not. 
However, one problem is that the relationship
between $R_{23}$ and 12+log(O/H) is double valued, with the transition
between the upper metal-rich branch and the lower metal-poor branch
occurring near 12+log(O/H)$\sim $8.4 (log$R_{23}$$ \sim $0.8).
Moreover, in many cases $R_{23}$ is
unavailable due to the limit of wavelength coverage or the poor quality
around one or some of the related lines. One can then use other various
metallicity-sensitive strong-line ratios to estimate metallicities of
galaxies, including 
[N~\textsc{ii}]$\lambda $6583/H$\alpha $, [O~\textsc{iii}]$\lambda
$5007/[N~\textsc{ii}]$ \lambda $6583, [N~\textsc{ii}]$\lambda
$6583/[O~\textsc{ii}]$\lambda $3727, [N~\textsc{ii}]$\lambda
$6583/[S~\textsc{ii}]$\lambda $$\lambda $6717,6731,
[S~\textsc{ii}]$\lambda $$\lambda $6717,6731/H$\alpha $, and
[O~\textsc{iii}]$\lambda $$\lambda $4959,5007/H$\beta $. Even when
$R_{23}$ is available, these ratios are also useful for breaking the
$R_{23}$ upper/lower branch degeneracy. We should notice that some of
these strong-line ratios are also sensitive to interstellar reddening
(e.g. N~\textsc{ii}]/[O~\textsc{ii}]) or depend on the ionization
parameter.

Some researchers have used $T_e$-based abundance data to
calibrate for some of these strong-line ratios, for
example, Denicol{\' o} et al. (2002) obtained a linear calibration of
[N~ \textsc{ii}]/H$\alpha $ to 12+log(O/H) from a combined sample of 108
metal-poor galaxies with [O~\textsc{iii}]4363 detected and 
128 metal-rich galaxies; Pettini \& Pagel (2004) studied
the calibration relations between [N~\textsc{ii}]/H$\alpha $,
([O~\textsc{iii}]/H$\beta $ )/([N~\textsc{ii}]/H$\alpha $) and
12+log(O/H) from 137 galaxies with $T_{e}$ -derived O/H abundances
(including six of them from photoionization models). P\'{e}rez-Montero
\& D\'{\i}az (2005) also studied some calibrations from the collected 367
emission-line objects on the basis of the $T_e$-based oxygen abundances,
but they did not derive analytic calibrating relations.
However, these observations are limited to
modest samples, and only apply to some of the strong-line
ratios.  The much larger consistent data set (several tens of thousands)
from the recently released SDSS database could provide much more
information about many of the strong-line ratio calibrations for
metallicities of galaxies.

KD02 used a combination of stellar population synthesis and
photoionization models to 
calibrate several strong-line abundance
 diagnostics.  In the KD02 models, strong line ratios are a function
 of both metallicity and ionization parameter.  The ionization
 parameter, $q$, is a measure of the number of ionizing photons
  per atom at the inner boundary of the nebula.  
 To determine
 both metallicity and ionization parameter, KD02 suggested a somewhat
 complicated iterative approach involving a number of line ratios.  We
 advocate a simpler approach in this study.  
 In Sect.\,4 we compare the observed line
 ratios of a large sample of star-forming galaxies selected from the SDSS-DR2
 to the predictions of the KD02 models and show that galaxies
 span a smaller range of $q$ than used by KD02.  We therefore ignore any
 explicit dependence of the line ratios on $q$, and use observational
 data to derive strong-line abundance calibrations which reflect the mean
 ionization parameter of the data at a given metallicity.

 The SDSS-DR2 database is very well suited for this kind of study.
 There are a large number of galaxies at any given value of oxygen
 abundance between about 8.4 and 9.3, which makes it ideal for
 measuring correlations between metallicity and various strong-line
 ratios and for quantifying the scatter. We make use of the emission
 line data and metallicities described in Brinchmann et al. (2004) and
 Tremonti et al. (2004) which have been made publicly available on
 the web\footnote{http://www.mpa-garching.mpg.de/SDSS/}. 
 The oxygen abundances presented in Tremonti et al. (2004) were
  derived using Bayesian techniques to compare multiple strong
  emission lines to a grid of photoionization models (Charlot et al. 2006).  
 We use the $R_{23}$ metallicity calibration of Tremonti et al. 
 to derive a fiducial oxygen abundance. 
  We then examine correlations
 between the oxygen abundances and the observed several 
 strong emission-line ratios:
 [N~\textsc{ii}]/H$\alpha $, [O~\textsc{iii}]/[N~\textsc{ii}],
 [N~\textsc{ii}]/[O~\textsc{ii}], [N~\textsc{ii }]/[S~\textsc{ii}],
 [S~\textsc{ii}]/H$\alpha $, and [O~\textsc{iii}]/H$\beta$, then  
 derive new strong-line metallicity calibrations where appropriate.  
 Our approach differs from the usual way in which 
 abundance indicators are calibrated: 
 1) using samples of galaxies with direct ($T_e$-based) abundances
 (e.g. Pagel et al. 1979; Pettini \& Pagel 2004); 
 2) using photoionization
 model grids (e.g. McGaugh 1991; Charlot \& Longhetti 2001; KD02). 
 Our calibrations are valid for high metallicity (12+log(O/H) $>$ 8.4)
 galaxies, where the $T_e$-based abundances are only available for
 a handful of objects (Bresolin et al. 2004, 2005; Garnett et
 al. 2004a,b). 
  We note that our calibrations
 may be subject to the same systematic uncertainty as the models.  For
 comparison, we use the $R_{23}$ calibration of McGaugh (1991) 
 (as Kobulnicky et al. 1999) to derive the fiducial metallicities as well.

We also estimate the log(N/O)
abundance ratios of these sample galaxies by using the algorithm given
by Thurston et al. (1996), including the estimated temperature in the
[N~ \textsc{ii}] emission regions. This will be the first such large
sample for the observed log(N/O) abundances, and will be useful for
understanding the \textquotedblleft primary" or 
\textquotedblleft secondary" origin of nitrogen. Specially, the
strength of the SDSS dataset is that there are a large number of
galaxies at any given value of oxygen abundance between about 8.4 and
9.3. This makes it ideal for quantifying the scatter in various
trends.

This paper is organized as follows: the sample selection and $R_{23}$
-derived oxygen abundances are described in Sect.2; Sect.3 shows the derived
calibrations of the linear and/or third-order polynomial fits from the
observational relations of 12+log(O/H) versus the various strong
emission-line ratios; these observational relations are compared with the
photoionization models of KD02 in Sect.4; The log(N/O) abundance ratios of
the sample galaxies are given in Sect.5; and we conclude in Sect.6.

\section{Sample selection, and R$_{23}$-derived oxygen abundances}

The data analyzed in this study are drawn from the SDSS-DR2 (Abazajian et 
al. 2004). These galaxies are part of the SDSS ``main" galaxy sample used 
for large-scale structure studies (Strauss et al. 2002). We select the 
sample galaxies having Petrosian r magnitudes in the range 14.5$<r<$17.77 
mag after correction for foreground Galactic extinction using the reddening 
maps of Schlegel et al. (1998), which yields 193,890 objects from the 
total of 261,054 objects.

We only consider galaxies whose metallicities have been estimated by
Tremonti et al. (2004), leaving 50,385 galaxies. They selected 
star-forming galaxies using the criteria given by Kauffmann et al. (2003b)
following the traditional line diagnostic diagram of [N~\textsc{ii}]/H$
\alpha $ vs. [O~\textsc{iii}]/H$\beta $ which has been used by Baldwin,
Phillips, Terlevich (1981, BPT), Veilleux \& Osterbrock (1987) and Kewley et
al. (2001) etc.

Tremonti et al. (2004) discussed the weak effect of the 
3{$^{\prime \prime }$} aperture of the SDSS spectroscopy 
on estimated metallicities of the sample
galaxies with redshifts $0.03<z<0.25$. Kewley et al. (2005) further
recommended that, to get reliable metallicities, redshifts $z>0.04$ are
required for the SDSS galaxies to ensure a covering fraction $>$20\% of the
galaxy light. Thus, we select the galaxies with redshifts $0.04<z<0.25$,
which then leaves 40,693 objects.

We only consider the objects for which fluxes have been measured for [O~
\textsc{ii}], H$\beta $, [O~\textsc{iii}], H$\alpha $, [N~\textsc{ii}] (also
[S~\textsc{ii}] in the study in Sects. 3.5, 3.6). 
It leaves 40,293 objects (39,919 if the [S~\textsc{ii}]6717,6731 are 
included to be considered as well).
Also, the sample objects
should have higher S/N ratios by requiring them to have lines of H$\beta $, H
$\alpha $, and [N~\textsc{ii}]6583 detected at greater than 5$\sigma $.
It leaves 39,029 objects.

The oxygen abundances of the sample galaxies were estimated using the $
R_{23}$ method. The emission-line fluxes must first be corrected for dust
extinction. The extinction correction of the sample galaxies are derived
using the Balmer line ratio H$\alpha $/H$\beta $: assuming case B
recombination, with a density of 100\thinspace cm$^{-3}$ and a temperature
of 10$^{4}$\thinspace K, and the predicted intrinsic ratio of H$\alpha $/H$
\beta $ is 2.86 (Osterbrock 1989), with the relation of 
\begin{equation}
(\frac{I_{H\alpha }}{I_{H\beta }})_{obs}  = 
(\frac{I_{H\alpha_0}}{I_{H\beta_0}})_{intr}10^{-c(f(H\alpha )-f(H\beta ))}.
\end{equation}
 Using the average interstellar extinction
law given by Osterbrock (1989), we have $f(\mathrm{H}\alpha )-f(\mathrm{H}
\beta )$=$-0.37$. Then, the extinction parameter $A_{V}$ is calculated
following Seaton (1979): $A_{V}=E(B-V)R=\frac{cR}{1.47}$ (mag). $R$= 3.1 is
the ratio of the total to the selective extinction at $V$. The emission-line
fluxes of the sample galaxies have been corrected for this extinction. For
the 311 data points with $A_{V}$$<$0, we assume their $A_{V}$$=$0 since
their intrinsic $H\alpha /H\beta $ may be lower than 2.86.

\begin{figure}[tbp]
\centering
\includegraphics[width=7.2cm]{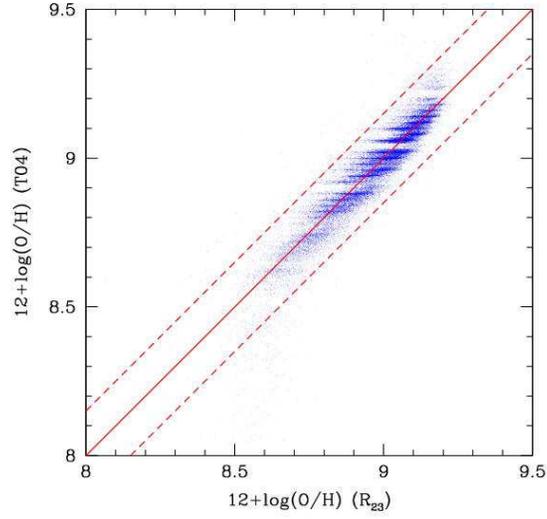}   
\caption{Comparison between our metallicity estimates from the
extinction-corrected $R_{23}$ parameters ($R_{23}$) and those 
Bayesian estimates from Tremonti et al. (2004) (T04) derived
using the model of Charlot et al. (2006). 
These two estimates are very consistent.
The solid line refers to the line of equality, and the $\pm$0.15\,dex
discrepancies from the solid line are also given as the two dashed lines. 
}
\label{fig1}
\end{figure}

Tremonti et al. (2004) used the approach outlined by Charlot et al. (2006), 
which consists of estimating metallicities statistically, based on 
simultaneous fits of all the most prominent emission lines ([O~\textsc{ii}],
H$\beta$, [O~\textsc{iii}], He~\textsc{i}, [O~\textsc{i}], H$\alpha$, 
[N~\textsc{ii}], [S~\textsc{ii}]) with a model designed  for the interpretation
of integrated galaxy spectra. Here, we use the  corresponding emission-line
fluxes to directly get dust  extinction-corrected $R_{23}$ values first,
and then adopt the calibration  given by Tremonti et al. (2004) (their eq.1:
$12+\mathrm{log(O/H)}=9.185-0.313x-0.264x^2-0.321x^3$, with  $x=\mathrm{log}
R_{23}$) to estimate the 12+log(O/H) abundances for our sample  galaxies,
which we denote as 12+log(O/H)$_{R_{23}}$. This calibration is only appropriate to the
metal-rich branch.  The consistency of these two estimates  is shown in 
Fig.~\ref{fig1}. The solid line is the line of  equality, and the two
dashed lines are $\pm$0.15\,dex discrepancies from the line of equality.
We notice that, for about 260 data points with
 $12+\mathrm{log(O/H)}
<8.5$ estimated by Tremonti et al. (2004), our
 $R_{23}$-derived abundances are higher than those of Tremonti.  All of
  these galaxies have -1.2$<$[N~\textsc{ii}]/H$\alpha $$<$-0.8 (see Fig.~\ref{fig3}a), 
  which implies that
  they are metal poor galaxies on the lower $R_{23}$ branch. Because the
  calibration of Tremonti et al. is only suitable for the upper
  branch, we select the galaxies having 8.4$<$12+log(O/H)$<$10 (39,006
  objects).  We also require the two metallicity estimates to be
  consistent to $\pm$0.15\,dex, which leaves a final sample of 38,478
  galaxies.

 If a histogram distribution of the 12+log(O/H)$
_{R_{23}}$ is presented, it will show a range of 8.4$<$$12+\mathrm{log(O/H)}$
$<$9.3 with a peak around 12+log(O/H)=9.0 and an increasing gradient-like
distribution following the increasing abundance from 8.4 to 9.1.
Among these selection criteria, only the 0.04 lower limit of redshift,
the 8.4$<$12+log(O/H)$<$10 and the 0.15\thinspace dex discrepancy were 
not considered by Tremonti et al. (2004), others are the same. 

\section{Deriving calibrations for oxygen abundances from the observations
of the SDSS galaxies}

We derive the analytic linear least squares and/or third-order polynomial fits
from the observed relations of 12+log(O/H)$_{R_{23}}$ 
versus \textquotedblleft strong-line" ratios of the SDSS sample galaxies: 
[N~\textsc{ii}]/H$\alpha $, [O~\textsc{iii}]/[N~\textsc{ii}], 
[N~\textsc{ii}]/[O~\textsc{ii}],
[N~\textsc{ii}]/[S~\textsc{ii}], [S~\textsc{ii}]/H$\alpha $, and 
[O~\textsc{iii}]/H$\beta $. To minimize the systematic uncertainty from
calibrations of $ R_{23}$ to O/H, we re-derive these calibrations by
using two other sets of oxygen abundance estimates replacing
12+log(O/H)$_{R_{23}}$: 12+log(O/H)$ _{K99}$, the $R_{23}$-derived
abundances by using the $R_{23}$-O/H calibration of Kobulnicky et al.
(1999, from McGaugh 1991), and 12+log(O/H)$ _{T04}$, the Bayesian
abundances given by Tremonti et al. (2004). All the calibration
coefficients and the rms derivations have been given in 
Table\thinspace1. 12+log(O/H)$_{K99}$ is a bit lower than the other two, less than $\sim
$0.1\thinspace dex lower. In this paper, we only plot the calibration
results of 12+log(O/H)$_{R_{23}}$ versus line ratios as representatives.

The derived calibrations in this paper are only appropriate for the
metal-rich galaxies (12+log(O/H)$>$8.4), because the metallicities of
these SDSS sample galaxies are in the range of 8.4$<$12+log(O/H)$<$9.3.
The corresponding ranges of the various related strong-line ratios will
be given in consequent subsections and Table\thinspace 1.
We omit the galaxies with lower metallicities,
12+log(O/H)$<$8.4, in calibrations, and leave them to future work.

\subsection{Aperture effects in the SDSS}

Aperture bias is clearly important when comparing metallicities measured in
the central regions of galaxies to \textquotedblleft global" quantities such
as the luminosity (Kewley et al. 2005). The main question here is if the
relation between the metallicity and the ionization parameter is the same in
nuclear and global spectra. To estimate the aperture effect on the
metallicity estimates, we compare our results about the SDSS galaxies with
those of the local Nearby Field Galaxy Survey (NFGS) galaxies from Jansen et
al. (2000a,b). Jansen et al. (2000b) have published the emission-line fluxes
from both the integrated and nuclear spectra of their galaxies. We
calculated their 12+log(O/H) vs. [N~\textsc{ii}]/H$\alpha $ relations from
their integrated and nuclear spectra, respectively, by using the same
method as we adopted for the SDSS galaxies.

\begin{figure}[tbp]
\centering
\includegraphics[width=7.2cm]{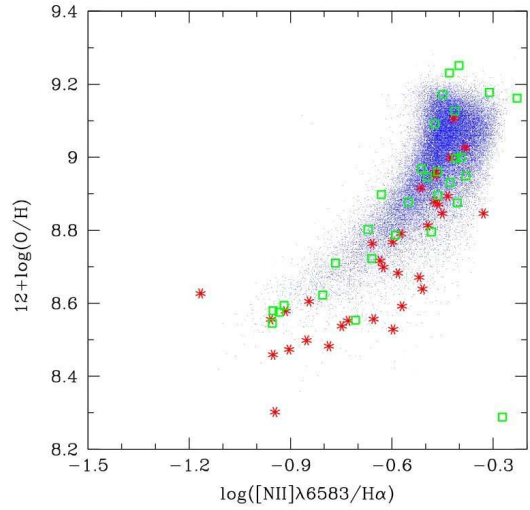}   
\caption{Comparison between our sample galaxies (the small blue points) and
the NFGS galaxies at local from Jansen et al. (2000a,b): ${the~ stars}$
refer to the results from the integrated spectra, and ${the~ open~
squares}$ refer to the results from the nuclear spectra (see text). }
\label{fig2}
\end{figure}

We select 38 galaxies from the sample of Jansen et al. (2000b)  following
the criteria:  
1) having the related fluxes for both the  integrated and
nuclear spectra available;  
2) confirming that they are star-forming galaxies by adopting the same
criterion of Kauffman et al. (2003b) as we used for the SDSS galaxies;  
3) using the same calculation method as for the SDSS galaxies, for
example, for estimating $A_V$ and deriving (O/H) abundances etc. 
Fig.~\ref{fig2} shows the comparison
with  the NFGS galaxies from the integrated spectra (the stars) and
the nuclear spectra (the open squares). It is clear that all the
data points from the  nuclear spectra follow the SDSS galaxies very well,
but some data points from the  integrated spectra show lower 12+log(O/H)
at a given [N~\textsc{ii}]/H$\alpha$ value, $\sim$0.1-0.2\,dex lower.
Fig.~\ref{fig2} may suggest a relatively lower ionization parameter for
the outskirt regions of galaxies, but this conclusion is limited by the 
small number of galaxies with integrated spectra.  Therefore, these
calibrations from SDSS galaxies may be more suitable to the  nuclear
spectra of galaxies.

\subsection{12+log(O/H) versus [N~\sc{ii}]/H$\protect\alpha$}

The [N~\textsc{ii}]/H$\alpha $ ratio is sometimes used to estimate metallicities of
galaxies since it is not affected very much by dust extinction due to their
close wavelength positions (see Pettini \& Pagel 2004). 
The near infrared spectroscopic instruments,
which were developed, can gather these two lines from the galaxies with
intermediate and high redshifts. 

Metallicity calibrations of log([N~\textsc{ii}]$\lambda
$6583/H$\alpha $) (the N2 index) can be found in
Storchi-Bergmann, Calzetti, \& Kinney (1994) and Raimann et al. (2000).
Recently, Denicol{\' o} et al. (2002) presented 
an analytical linear formula by fitting the
relation of a combined sample of 108 metal-poor galaxies (their O/H
abundances were estimated from $T_e$) and 128 metal-rich galaxies 
(their O/H abundances were estimated from $R_{23}$  or $S_{23}$ (=
([S~\textsc{ii}]6717,6731+[S~\textsc{iii}]9069,9531)/H$\beta$):
 $12+\log \mathrm{(O/H)}=9.12+0.73\times N2$. 
The data of P\'{e}rez-Montero \& D\'{\i}az (2005) are
generally consistent with this calibration. Pettini \& Pagel (2004) obtained
a similar linear fit for a sample of 137 galaxies. Most 
of their samples have 7.0$<$
12+log(O/H)$<$8.6 based on $T_{e}$ measurements, and six (four are
metal-rich) have oxygen abundances obtained from a detailed
photoionization model. The linear fit given by Pettini \& Pagel (2004) is: $
12+\log \mathrm{(O/H)}=8.90+0.57\times N2,$ and another third-order
polynomial fit with slight revision is: $12+\log \mathrm{(O/H)}
=9.37+2.03\times N2+1.26\times N2^{2}+0.32\times N2^{3}$ (valid in the range
of $-2.5<N2<-0.3$).

\begin{figure*}[tbp]
\centering
\includegraphics[width=16.8cm]{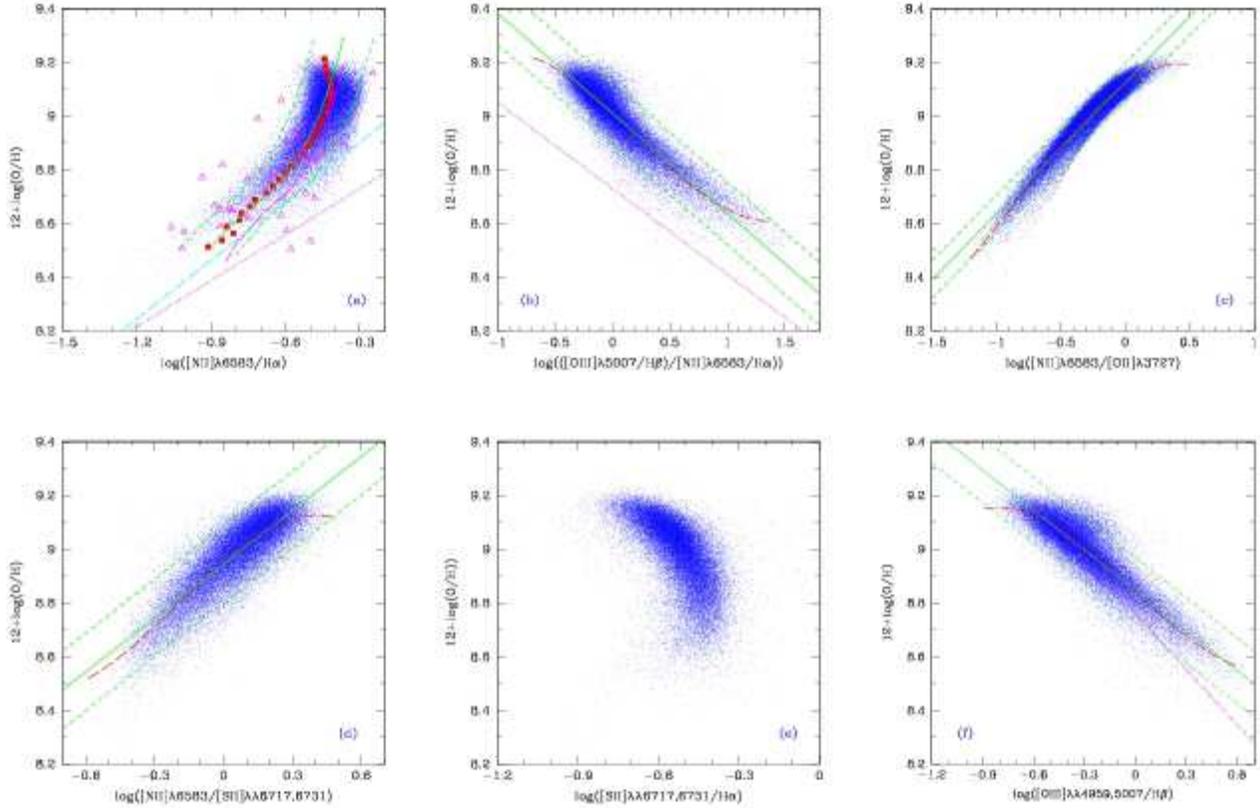}   
\caption{ {\protect\footnotesize Analytic calibration relations between
12+log(O/H) and several metallicity-sensitive strong-line ratios based on
the observed relations of the SDSS sample galaxies ($the$ $small~ blue~
points$):
(\textbf{a.}) versus N2 index (without reddening correction for 
[N\textsc{ii}]/H$\protect\alpha$). 
The large squares represent the 29 median values
in bins of 0.025 dex in 12+log(O/H) within 12+log(O/H) from 8.5 to 9.3.
The open triangles refer the local H~\textsc{ii} galaxies from 
Denicol{\' o} et al. (2002) (from Terlevich et al. 1991). 
The dot-long-dashed line represents the
calibration with 12+log(O/H)$_{K99}$. 
(\textbf{b.}) versus O3N2 index relations. 
(\textbf{c.}) versus log([N~\textsc{ii}]/[O~\textsc{ii}]). 
(\textbf{d.}) versus log([N~\textsc{ii}]/[S~\textsc{ii}]) (without reddening
correction for [N\textsc{ii}]/[S~\textsc{ii}]). 
(\textbf{e.}) versus log([S~\textsc{ii}]/H$\protect\alpha$). 
(\textbf{f.}) versus log([O~\textsc{iii}]/H$\protect\beta$) ratios. 
The solid lines 
refer to the fits to the data (see text), and the two dashed-lines
show the discrepancy of 2$\protect\sigma$; the long-dashed lines in 
\textbf{Figs.b,c,d,f} refer
to the three-order polynomial fits for the observational relations. 
The dotted
lines in \textbf{Figs.a,b} refer to the linear least-squares fit given by Pettini \& Pagel
(2004), the long-dashed line in \textbf{Fig.a} refers
to the result of Denicol{\' o} et al. (2002). 
The dot-long-dashed line in \textbf{Fig.f} refers to the
linear least-squares fit given by Vacca \& Conti (1992). 
All the fitting coefficients for the SDSS galaxies and the rms derivations
 have been given in Table\,1 following the format of Eq.(2) and Eq.(1).}}
\label{fig3}
\end{figure*}

Here we present the relation of N2 index to (O/H)$_{R_{23}}$ for the 38,478
SDSS star-forming galaxies, and derive a corresponding calibration from
these metal-rich galaxies. Fig.~\ref{fig3}a shows our results for
12+log(O/H)$_{R_{23}}$ vs. N2 index relations (the small blue points, without
reddening-correction for [N~\textsc{ii}]/H$\alpha $ ratios). The large red
squares represent the 29 median N2 values in bins of 0.025 dex in
12+log(O/H) within the range of 12+log(O/H)=8.5 to 9.3. The 12+log(O/H)
shows a clear increasing trend following the increase of the N2 index, up to
12+log(O/H)$\sim $ 9.0, and then, 
the galaxies having 12+log(O/H)$>$ 9.0 show a
slightly decreasing N2 index.  
This trend can be understood as follows: when the
secondary production of nitrogen dominates, at somewhat higher metallicity,
the [N~\textsc{ii}]/H$\alpha $ line ratio continues to increase, despite the
decreasing electron temperature; eventually, at still higher metallicities,
nitrogen becomes the dominant coolant in the nebula, and the electron
temperature falls sufficiently to ensure that the nitrogen line weakens with
increasing metallicity (KD02). This is the first large sample to show this
turnover relation between N2 and 12+log(O/H) at the very metal-rich region.
An analytical third-order polynomial fit to the median-value points can be
given in Fig.~\ref{fig3}a as the solid line: 
\begin{eqnarray}
12+\mathrm{log(O/H)} &=&a0+a1\times N2+a2\times N2^{2}  \nonumber \\
&+&a3\times N2^{3}.
\end{eqnarray}
The fitting coefficients a0, a1, a2 and a3 are given in Table\thinspace 1, 
as well the rms derivation ($\sigma $=0.058\thinspace dex) 
for the data points from the fitting relation.
This fit is obtained from the data points with $-1.2<N2<-0.2$ and 8.4$<$
12+log(O/H)$<$9.3. 
The two dashed lines describe the discrepancy of 2$\sigma $. A linear fit
does not work well for these relations.

However, we should notice that this calibration could result in large
uncertainty when 12+log(O/H)$>$9.0 because of the turnover of N2 index
versus O/H in the very metal-rich region. Thus, we re-derive a third-order
polynomial fit for this calibration for the part of sample galaxies having
8.4$<$12+log(O/H)$<$9.0. We also re-derive the
calibrations of the N2 index to 12+log(O/H)$_{K99}$ (the dot-long-dashed
line in magenta color in Fig.~\ref{fig3}a) and 12+log(O/H)$_{T04}$. All the
fitting coefficients and the rms derivation $\sigma $ values are given in
Table\thinspace 1.

In Fig.~\ref{fig3}a we compare our calibration of [N~\textsc{ii}]/H$\alpha$ vs. 
12+log(O/H) with other two calibrations published recently, as shown by the 
long-dashed and the dotted lines in Fig.~\ref{fig3}a. The long-dashed line 
represents the linear least squares fit given by Denicol{\' o} et al. 
(2002) for the whole sample of their 236 galaxies.
The  dotted line
refers to the linear least squares fit obtained by Pettini \&  Pagel (2004)
from their 137 sample galaxies.  Both of these two calibrations show relatively
lower O/H abundance  compared to ours at a given N2 index, especially at the
very  metal-rich part.

The difference from Pettini \& Pagel (2004) is easily understood since they
only select metal-poor galaxies with $T_{e}$ measurements, except four
metal-rich and two metal-poor ones. The fit for these metal-poor galaxies
can be directly extrapolated to the metal-rich region and results in a lower
O/H at a given [N~\textsc{ii}]/H$\alpha $ value there. This may be related
to the suggestion of Kennicutt et al. (2003), who pointed out that the
abundance estimates from $T_{e}$ method are systematically lower by
0.2-0.5\thinspace dex than those derived from the strong-line
\textquotedblleft empirical" abundance indicators, when the latter are
calibrated with theoretical photoionization models.

Denicol{\' o} et al. (2002) analyzed their metal-poor and metal-rich
galaxies together and used the calibration of Kobulnicky et al. (1999) for
converting $R_{23}$ to O/H for most of the metal-rich galaxies. Their fig.1
shows that most of the metal-rich samples indeed lie on the left-top side of
the fitting line. 
If only their metal-rich galaxies were considered to derive a calibration,
it will predict a higher oxygen abundance  
at a given [N~\textsc{ii}]/H$\alpha $ ratio than that they  
derived from the combined sample. 
To be clear, we re-derive the [N~\textsc{ii}]/H$
\alpha $ and 12+log(O/H) values for their metal-rich sample galaxies taken
from Terlevich et al. (1991), in which the fluxes of the related lines were
given as a table. The open triangles in Fig.~\ref{fig3}a present the H~
\textsc{ii} galaxies from Terlevich et al. (1991) by using the same
calibration as ours for the SDSS galaxies. The two samples show consistency.
And the data of Terlevich et al. (1991) show more scatter, perhaps it comes
from the different slit widths and resolutions used in different runs with
different telescopes.

\subsection{12+log(O/H) versus [O~\sc{iii}]/[N~\sc{ii}]}

Alloin et al. (1979) firstly introduced the quantity  
$O3N2= \log $\{([O~\textsc{iii}]~$\lambda 5007$/H$\beta$)$/
$([N~\textsc{ii}]~$\lambda 6583$/H$\alpha$)\}\footnote{
This definition is  slightly different from the original one proposed by
Alloin et al. (1979)  who included both [O~\textsc{iii}] doublet lines in the
numerator of the first ratio.}.  Recently, Pettini \& Pagel (2004) presented
one such calibration  by doing linear least squares fitting for a sample of
65 (from the 137)  galaxies in the range of $-1 < O3N2 < 1.9$. They obtained
a calibration of  $ 12 + \log \mathrm{(O/H)} = 8.73 - 0.32 \times O3N2.  $

We can obtain such a calibration from the large sample of SDSS galaxies.
Fig.~\ref{fig3}b shows the $R_{23}$-derived 12+log(O/H) versus O3N2
relations for our sample galaxies. The solid line gives a linear least
squares fit: 
\begin{equation}
12+\log \mathrm{(O/H)}=b0+b1\times O3N2,
\end{equation}
with the coefficients b0 and b1 given in Table\thinspace 1, as
well as the rms derivation. The two dashed lines show the discrepancy of 
2$\sigma $. The dotted line refers to the fit of Pettini \& Pagel (2004)
for their 65 metal-poor galaxies, which gives lower O/H than that of our
samples at a given O3N2 value; about 0.1\thinspace dex lower. The reason may
be that the galaxies with $T_{e}$-abundances were mainly
used in their study.

To get a better fit for the details of the relation, a third-order
polynomial fit is obtained and shown as the long-dashed line in 
Fig.~\ref{fig3}b, which revises the linear equation slightly. The corresponding
fitting coefficients of a0, a1, a2 and a3 are given in Table\thinspace 1
with the same meaning as Eq.(1), as well as the rms derivation. These linear
and three-order polynomical calibrations were obtained from the sample
galaxies with 8.4$<$ 12+log(O/H) $<$9.3 and -0.7 $<$O3N2$<$ 1.6. The obvious
advantages of using the [N~\textsc{ii}] and [O~\textsc{iii}] lines is that
they are unaffected by absorption lines originating from the underlying
stellar populations, they lie close to Balmer lines that can be used to
eliminate errors due to dust reddening, and they both are strong and easily
observable in the optical band.

\subsection{12+log(O/H) versus [N~\sc{ii}]/[O~\sc{ii}]}

We can derive the calibration of [N~\textsc{ii}]/[O~\textsc{ii}] ratio to
12+log(O/H) from the large sample of the SDSS galaxies. Fig.~\ref{fig3}c
presents these data points in the relation of 12+log(O/H) versus 
[N~\textsc{ii}]/[O~\textsc{ii}]. A linear least squares fit reproduces the data
distribution well, as shown by the solid line. The two dashed lines
show 2$\sigma $ discrepancies. A third-order polynomial fit will revise
the fitting equation slightly. The coefficients of the two fits and the rms
derivation $\sigma $ of the data to the line are given in Table\thinspace 1,
as well as the appropriate ranges of 12+log(O/H) and 
log([N~\textsc{ii}]/[O~\textsc{ii}]). 
The fitting formulas follow the same format as Eq.(2) and Eq.(1).

Fig.~\ref{fig3}c shows that the [N~\textsc{ii}]/[O~\textsc{ii}] ratios
obviously increase with the increasing metallicities, which can be explained
by the predominant secondary enrichment of [N~\textsc{ii}] above metallicities
of 12+log(O/H)$>$8.6 (Alloin et al. 1979; KD02), and the stronger decrease
in the number of collisional excitations of the blue [O~\textsc{ii}] lines
relative to the lower energy [N~\textsc{ii}] lines due to the low electron
temperature in high metallicity environments. 

\subsection{12+log(O/H) versus [N~\sc{ii}]/[S~\sc{ii}]}

We can derive the calibration of  [N~\textsc{ii}]6583/[S~\textsc{ii}]6717,
6731 to 12+log(O/H) by using the  large sample of SDSS galaxies.  
Fig.~\ref{fig3}d shows these observed results,  which can be explained by a
linear least squares fit and is  given as the solid line.  The two dashed
lines show the 2$\sigma$ discrepancy.  To get a more detailed fit, a
third-order polynomial fit  has been obtained (the long-dashed line in 
Fig.~\ref{fig3}d),  which revises the relation slightly.  
The coefficients of the
two fits and the rms derivations are given in  Table\,1, as well the
appropriate ranges of 12+log(O/H) and  log([N~\textsc{ii}]/[S~\textsc{ii}]).
The fitting formulas follow the same format as Eq.(2) and Eq.(1).

It shows that the 12 + log(O/H) abundances of the sample  galaxies increase
with their  log([N~\textsc{ii}]/[S~\textsc{ii}]) ratios. The reason may be
that at high metallicity, nitrogen is a secondary nucleosynthesis  element
and sulphur is a primary process element (KD02).  Both lines have similar 
excitation potential since they are close to each other in wavelength.

\subsection{12+log(O/H) versus [S~\sc{ii}]/H$\protect\alpha$}

Although [S~\textsc{ii}]/H$\alpha$ ratio can be easily derived from [N~
\textsc{ii}]/H$\alpha$  and [N~\textsc{ii}]/[S~\textsc{ii}] ratios, it is
still interesting to directly present the  [S~\textsc{ii}]/H$\alpha$ to
12+log(O/H) relations for this large sample of  SDSS galaxies.
Fig.~\ref{fig3}e shows that these data points show a double-valued
distribution:  because of the turnover around 12+log(O/H)$\sim$8.9-9.0, 
there will be double-valued oxygen abundances at a  given [S~\textsc{ii}]/H$
\alpha$ ratio.  Thus, it is difficult to fit a  simple and useful analytic
formula from these observational data.

We should notice that the [S~\textsc{ii}]/H$\alpha $ ratio has never been
suggested or used as a metallicity indicator, because it is far more
sensitive to ionization (see Sect.4) than to metallicity, and it results in
double-valued (O/H). If it is able to measure the [S~\textsc{ii}] and H$
\alpha $ lines, it is likely that we have also obtained [N~\textsc{ii}], and
then we could use the [N~\textsc{ii}]/H$\alpha $ calibration.

\subsection{12+log(O/H) versus [O~\sc{iii}]/H$\protect\beta$}

The [O~\textsc{iii}]4959,5007/H$\beta $ ratios can also be used to estimate
the metallicities of galaxies. This was called \textquotedblleft $R_{3}$
method" (Edmunds \& Pagel 1984), and was shown as: $R_{3}=1.35\times
(I_{[OIII]5007}/H\beta ),$ and then, $\mathrm{log(O/H)}=-0.69\times \mathrm{
log}R_{3}-3.24,$ with a range of -0.6 $\leq $ log$R_{3}$ $\leq $ 1.0 (Vacca
\& Conti 1992).

Here we present a new calibration for this ratio vs. O/H derived from the
SDSS sample  galaxies (Fig.~\ref{fig3}f).  The linear least squares and the
third-order  polynomial fits are obtained, and given as the solid and
the long-dashed lines in Fig.~\ref{fig3}f, respectively. The
coefficients of  the two fits and the rms derivations are given in Table\,1
following Eq.(2) and Eq.(1),  as well the appropriate ranges of 12+log(O/H) and
log$R_3$.  The two dashed lines show the 2$\sigma$ discrepancy.

The observated results of this large sample of SDSS star-forming
galaxies and the derived analytic calibrations  show that the higher 
[N~\textsc{ii}]/H$\alpha$,  [N~\textsc{ii}]/[O~\textsc{ii}] and 
[N~\textsc{ii}]/[S~\textsc{ii}],  and the lower [O~\textsc{iii}]/H$\beta$,  
([O~\textsc{iii}]/H$\beta$)/([N~\textsc{ii}]/H$\alpha$) ratios will result in the higher
12+log(O/H) abundances generally. The rms derivations of the observational
data relative to the calibration relations are in 0.032-0.078\,dex.  
[N~\textsc{ii}]/[O~\textsc{ii}] has the lowest rms derivation. These
calibrations from strong-line ratios to metallicities can be used to calibrate
oxygen abundances of galaxies in future studies.

\section{Comparison with the photoionization models}

\begin{figure*}[tbp]
\centering
\includegraphics[width=16.8cm]{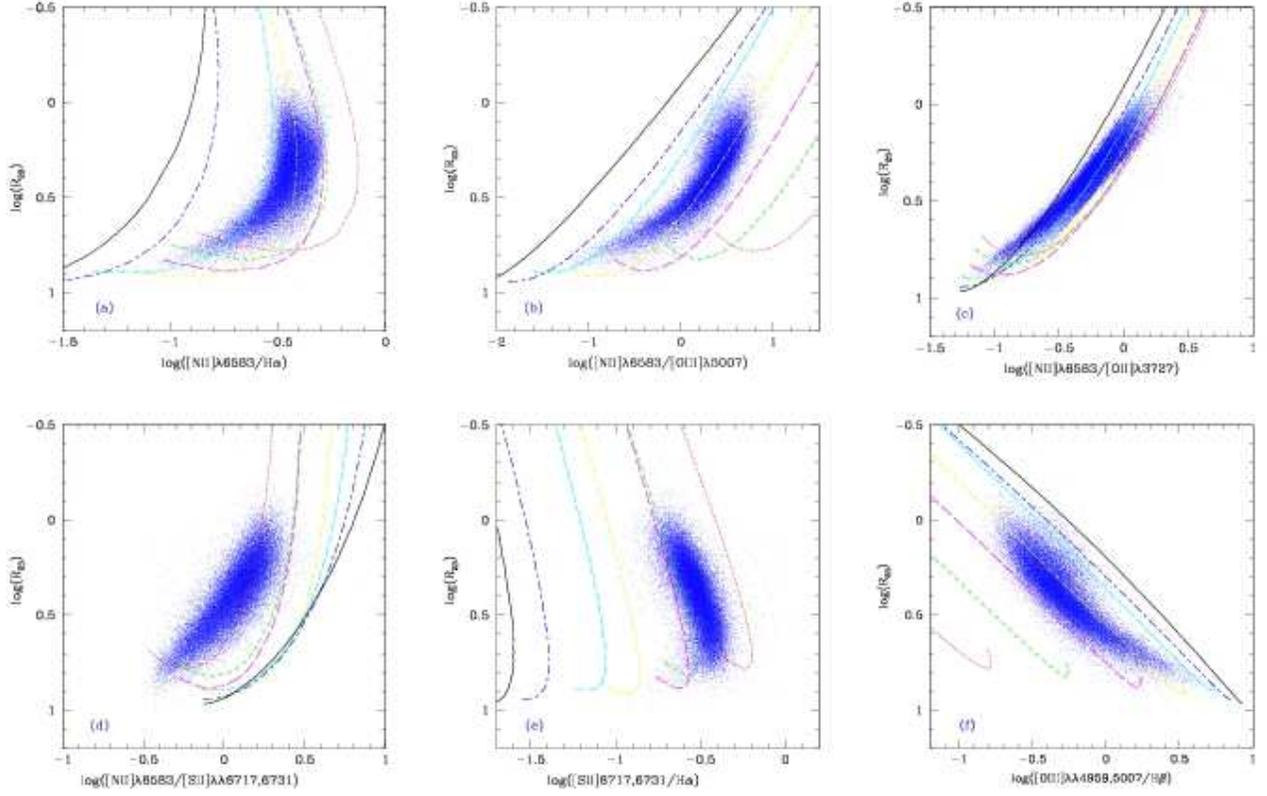}   
\caption{ Comparison between our SDSS sample galaxies ($the$ $small~ blue~
points$) and the photoionization model results of KD02 ($the~ lines$)
for the relations of log($R_{23}$) versus several strong-line ratios: 
\textbf{(a.)} log([N~\textsc{ii}]/H$\protect\alpha$);
\textbf{(b.)} log([N~\textsc{ii}]/[O~\textsc{iii}]); 
\textbf{(c.)} log([N~\textsc{ii}]/[O~\textsc{ii}]); 
\textbf{(d.)} log([N~\textsc{ii}]/[S~\textsc{ii}]); 
\textbf{(e.)} log([S~\textsc{ii}]/H$\protect\alpha$); 
\textbf{(f.)} log([O~\textsc{iii}]/H$\protect\beta$). 
The lines represent the model results of KD02 with  
$q=5\times 10^{6}$, $1\times 10^{7}$, $2\times 10^{7}$,
$4\times 10^{7}$, $8\times 10^{7}$, $1.5\times 10^{8}$, and 
$3\times 10^{8}$ cm\,s$^{-1}$, shown as 
from the red dotted line to the black solid line, respectively.
 The seven lines in {\bf Fig.f} were deduced from
other three line ratios, i.e. log([N~\textsc{ii}]/[O~\textsc{iii}]), log([N~
\textsc{ii}]/[O~\textsc{ii}]) and log($R_{23}$), since there are no direct
log([O~\textsc{iii}]/H$\protect\beta$) calibration relation given in KD02. }
\label{fig4}
\end{figure*}

\begin{figure}
    \centering
    \includegraphics[width=7.8cm]{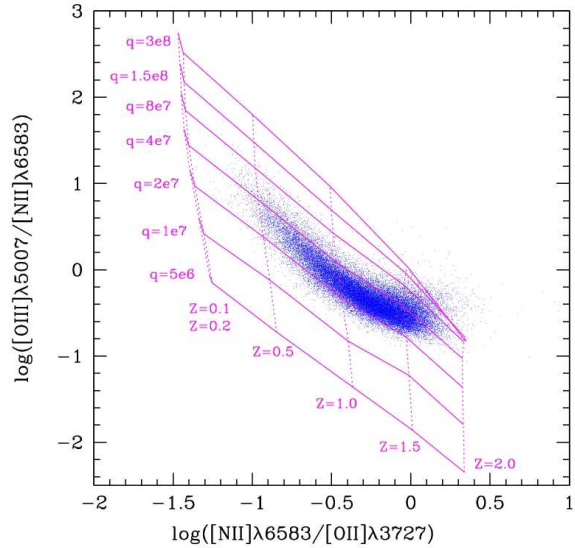}
       \caption{Comparison between the observed results of the
       SDSS sample galaxies
       and the photoionization models of Kewley et al. (2001)
       following fig.7 of Dopita et al. (2000).
         The solid lines
 refer to constant ionization parameters, with the grids of
 $q=5\times 10^6$, $1\times 10^7$, $2\times 10^7$,
 $4\times 10^7$, $8\times 10^7$, $1.5\times 10^8$, $3\times 10^8$
 cm\,s$^{-1}$; and the dotted lines
 refer to constant metallicity, with the grids of
 $Z=$0.1, 0.2, 0.5, 1.0, 1.5 and $2.0Z_{\odot}$.
                 }
    \label{fig5new}
    \end{figure}

 Our strong-line calibrations are in good qualitative agreement with
 those of KD02.  For example, both calibrations show a turnover of
 [N~\textsc{ii}]/H$\alpha $ near 12+log(O/H)$\sim $9.0.  
 One major difference is that KD02
 provided strong-line calibrations for seven different values of the
 ionization parameter, whereas we assume no explicit ionization
 parameter dependence.  Our approach will be less accurate in
 some cases, but it is necessary when only a few emission lines can be
 measured and there is too little information to determine both
 ionization parameter and metallicity.  It is also unclear that the full
 range of ionization parameters used in the KD02 models are required by
 the data. We explore this in Fig.~\ref{fig4}, where we plot the relations
 between $R_{23}$ and other strong line ratios in the SDSS data and overlay
 the KD02 model grids. Fig.~\ref{fig5new} also gives the similar information.

In Figs.~\ref{fig4}a-f, the y-axis log($R_{23}$) is
plotted from the lower to the higher values, which corresponds to an
increasing trend of metallicity for the galaxies in the metal-rich branch.
The seven lines in each plot represent the model results of KD02 for 
ionization parameters $q=5\times 10^{6}$, $1\times 10^{7}$, 
$2\times 10^{7}$, $4\times 10^{7}$, $8\times 10^{7}$, $1.5\times 10^{8}$, and 
$3\times 10^{8}$ cm\thinspace s$^{-1}$, respectively. From the dotted
line to the solid line, the ionization parameters increase orderly. The lines
in Figs.~\ref{fig4}a-f start from 12+log(O/H)=8.4 for the metal-rich branch,
and the turn-over trend of the model grids at the low-metallicity 
end corresponds
to the turnover of log($R_{23}$) itself from the metal-rich to the metal-poor
branches. 

--- [N~\textsc{ii}]/H$\alpha $ $~~~$ Fig.~\ref{fig4}a shows the comparison
for log($R_{23}$) versus log([N~\textsc{ii}]/H$\alpha $). It shows that
these observed points are reasonably consistent with the model results of
KD02. The increasing trend of [N~\textsc{ii}]/H$\alpha $ follow the
decreasing log($R_{23}$)and then the turnover from about log($R_{23}$)$\sim $
0.5 correspond to the increasing and then the turnover of 
[N~\textsc{ii}]/H$\alpha $ following the increasing 12+log(O/H) abundances as 
shown in Fig.~\ref{fig3}a. However, [N~\textsc{ii}]/H$\alpha $ is sensitive to the
ionization parameter. It seems that  $q=1.5\times
10^{8}$ and $3\times 10^{8}$ cm\thinspace s$^{-1}$
are too high and $q=5\times 10^{6}$ cm\thinspace s$^{-1}$ is too low to
explain the distribution of these star-forming galaxies. Most of the sample
galaxies fall in the range of ionization parameter $q$ from $1\times 10^{7}$
to $8\times 10^{7}$ cm\thinspace s$^{-1}$, which is narrower than the range
proposed by the models.

--- [N~\textsc{ii}]/[O~\textsc{iii}] $~~~$ Fig.~\ref{fig4}b shows the
comparison between the observed relations of log($R_{23}$) versus 
log([N~\textsc{ii}]/[O~\textsc{iii}]) with the photoionization model results of
KD02. It shows that the observed relations are consistent with the model
results: the log([N~\textsc{ii}]/[O~\textsc{iii}]) ratios of these galaxies
increase following the decreasing log($R_{23}$) (the increasing
metallicities). It also shows that, most of the data points distribute in a
range of ionization parameter $q$ from $q=1\times 10^{7}$ to $8\times 10^{7}$
cm\thinspace s$^{-1}$, which is similar to log([N~\textsc{ii}]/H$\alpha $).
[N~\textsc{ii}]/[O~\textsc{iii}] is sensitive to the ionization parameter.

---[N~\textsc{ii}]/[O~\textsc{ii}] $~~~$ Fig.~\ref{fig4}c shows the
comparison between the observed results and the photoionization model
results of KD02 for the relations of log($R_{23}$) versus 
log([N~\textsc{ii}]/[O~\textsc{ii}]). It shows that the 
[N~\textsc{ii}]/[O~\textsc{ii}] ratios of the sample galaxies increase 
following the decreasing log($R_{23}$) (the
increasing metallicities), which is very consistent with the general trend
of the model results. As KD02 mentioned, [N~\textsc{ii}]/[O~\textsc{ii}] is
almost independent of the ionization parameter. Our large set of data points
shows a very narrow distribution in the log($R_{23}$) vs. 
[N~\textsc{ii}]/[O~\textsc{ii}] relation, which is very consistent with 
KD02's model results. However, [N~\textsc{ii}]/[O~\textsc{ii}] ratios are 
affected strongly by dust extinction inside the galaxies. 
We will discuss this in Fig.~\ref{fig7} 
and in the last paragraph of this section.

--- [N~\textsc{ii}]/[S~\textsc{ii}] $~~~$ Fig.~\ref{fig4}d shows the
comparison between the observed results and the photoionization model
results of KD02 for the relations of log($R_{23}$) versus 
log([N~\textsc{ii}]/[S~\textsc{ii}]). It shows that the [N~\textsc{ii}]/[S~\textsc{ii}] 
ratios of the sample galaxies increase with the decreasing log($R_{23}$) (the
increasing metallicities), which is consistent with the general trend of the
models. However, the fact is that these SDSS data points show relatively lower
log($R_{23}$) (higher metallicities) than the model predicts at a given 
[N~\textsc{ii}]/[S~\textsc{ii}] ratio, about 0.2\thinspace dex discrepancy. One
of the reasons may be that the amount of S going into [S~\textsc{iii}]
compared to [S~\textsc{ii}] is not well understood, so this could introduce
some uncertainties into the photoionization models (see KD02). KD02
commented that their [N~\textsc{ii}]/[S~\textsc{ii}] technique will
systematically underestimate the abundance by $\sim $0.2\thinspace dex 
compared to the average of the comparison techniques.
The comparison of models and data shows that the models fail 
to reproduce the strengths of the [S~\textsc{ii}] lines, so 
we cannot use the models to infer abundances or ionization parameters from 
[N~\textsc{ii}]/[S~\textsc{ii}], as well [S~\textsc{ii}]/H$\alpha$ (see below).
These highlight the need for empirical calibration.

--- [S~\textsc{ii}]/H$\alpha $ $~~~$ Fig.~\ref{fig4}e shows the comparison
between the observed results and the photoionization model results of KD02
for the relations of log($R_{23}$) versus log([S~\textsc{ii}]/H$\alpha $).
There is a weak correlation between these two parameters, meaning that
log([S~\textsc{ii}]/H$\alpha $) decreases following the decreasing log($
R_{23}$) with a very weak turnover around log($R_{23}$)$\sim $0.5. For
the [S~\textsc{ii}]/H$\alpha $ calibration, the SDSS data points 
agree with a relatively higher model $q$ value than in the 
[N~\textsc{ii}]/[S~\textsc{ii}] case.

--- [O~\textsc{iii}]/H$\beta $ $~~~$ Fig.~\ref{fig4}f shows the comparison
between the observed results and the photoionization model results of KD02
for the relations of log($R_{23}$) versus log([O~\textsc{iii}]/H$\beta $).
The model results of [O~\textsc{iii}]/H$\beta $ from KD02 were deduced using
their three other ratio parameters, i.e. log([N~\textsc{ii}]/[O~\textsc{iii}]),
log([N~\textsc{ii}]/[O~\textsc{ii}]) and log($R_{23}$), since KD02 did not
give a direct formula for this calibration. These SDSS sample galaxies
follow the trend of the model results well, meaning the 
log([O~\textsc{iii}]4959,5007/H$\beta $) ratios decrease 
following the decreasing log($R_{23}$) (the increasing metallicities). 
This ratio depends strongly on the
ionization parameter. Most of the data points are in a range of $q=1\times
10^{7}$ cm\thinspace s$^{-1}$ to $8\times 10^{7}$ cm\thinspace s$^{-1}$,
except in the very metal-rich region with log($R_{23}$)$<$0.2. \newline

Figs.~\ref{fig4}a,b,f show that these star-forming galaxies have a
relatively narrower range of ionization parameter $q$ than the range
proposed by the photoionization models of KD02. The actual range of $q$ of
these galaxies is $1\times 10^{7}$$-$$8\times 10^{7}$ cm\thinspace s$^{-1}$,
narrower than the range of models, $q=5\times 10^{6}$$-$$3\times 10^{8}$
cm\thinspace s$^{-1}$.  We do not
consider Figs.~\ref{fig4}c,d,e for this range since 
[N~\textsc{ii}]/[O~\textsc{ii}] is insensitive to $q$ and the models 
provide a poor match to [N~\textsc{ii}]/[S~\textsc{ii}] 
and [N~\textsc{ii}]/H$\alpha $ (KD02).
To show more clearly the range of $q$ of these sample galaxies,
we compare their observed results with the model grids in Fig.~\ref{fig5new}
for the [O~\textsc{iii}]/[N~\textsc{ii}] versus [N~\textsc{ii}]/[O~\textsc{ii}]
relations following Dopita et al. (2000) (their fig.7).
The model grids are taken from Kewley et al. (2001)\footnote{Thanks to 
 Lisa Kewley for putting their model grids on the
 web: http://www.ifa.hawaii.edu/$\sim$kewley/Mappings/},  
 their instantaneous zero-age starburst models based on the PEGASE
 spectral energy distribution (SED),
 which are the most appropriate to use to model H~{\sc ii} regions (KD02),
 and show information about both metallicity and $q$.
  This diagram shows that these nearby
 star-forming galaxies have
 metallicities $Z=0.2-2.0Z_{\odot}$, and the actual
 ionization level of their ionized gas
 have ionization parameters in the range of
 $q=1\times 10^7$ to $8\times 10^7$cm\,s$^{-1}$ with some exceptions
 at the very metal-rich end.

Indeed, the parameter $q$ is related to the ionization parameter  $U$ by
dividing the speed of light $c$:  $U=q/c$, which is more commonly used 
(e.g. McGaugh 1991).  Therefore, the ionization parameters of KD02's models
correspond to a range from log$U$=$-$3.78 to log$U$=$-$2.  Our Figs.~\ref
{fig4}a,b,f show that the actual range of ionization parameters  of these
SDSS sample galaxies are from about log$U$=$-$3.5 to $-$2.5.  
This range is also consistent with what found by Dopita et
al. (2000) for the  extragalactic H~\textsc{ii} regions.

\begin{figure}[tbp]
\centering
\includegraphics[width=7.8cm]{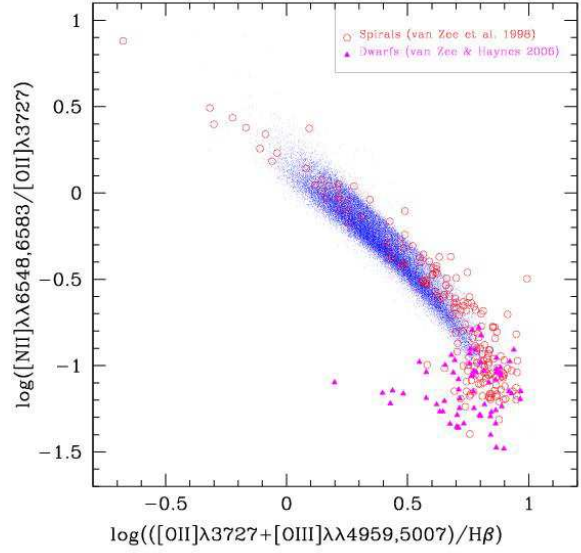}  
\caption{Comparison between our sample galaxies and the spiral galaxies form
van Zee et al. (1998) and the dwarf irregulars from van Zee \& Haynes
(2006). }
\label{fig5}
\end{figure}

We also compare these SDSS star-forming galaxies with the dwarfs from van
Zee \& Haynes (2006, the filled triangles) and local spirals from
van Zee et al. (1998, the open circles) in Fig.~\ref{fig5} on the basis
of the relation of [N~\textsc{ii}]/[O~\textsc{ii}] versus log$R_{23}$.
Although [N~\textsc{ii}]/[O~\textsc{ii}] is insensitive to ionization
parameters, log$R_{23}$ is, so this plot can help to understand the
evolutionary status of these sample galaxies. It is clear that these SDSS
galaxies follow the metal-rich spirals well, 
but the dwarfs 
  probe a metallicity range not covered by the SDSS galaxies$-$ 
  only very few dwarfs have 12+log(O/H)$>$8.4. 

\begin{figure}[tbp]
\centering
\includegraphics[width=7.8cm]{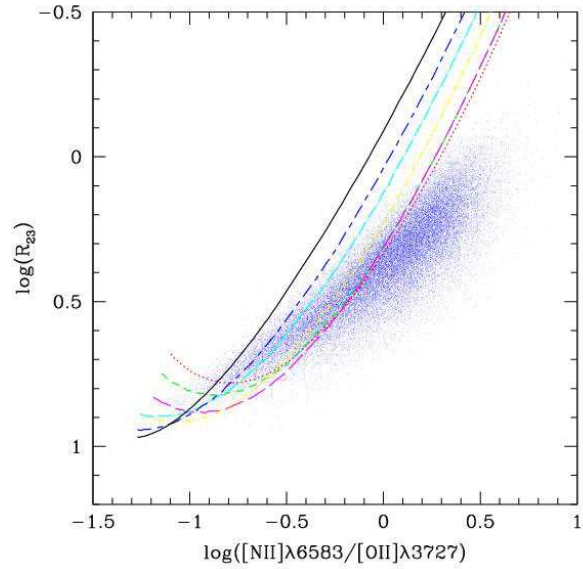}   
\caption{Comparison the log($R_{23}$) versus log([N~\textsc{ii}]/[O~\textsc{ii}]) 
relations between our sample galaxies without correcting dust extinction
for [N~\textsc{ii}]/[O~\textsc{ii}] and the model results of KD02. The
comparison of this plot with Fig.~\protect\ref{fig4}c shows that the dust
extinction strongly affect the [N~\textsc{ii}]/[O~\textsc{ii}] ratios and
their scatter, hence the calibrated metallicities. }
\label{fig6}
\end{figure}

We should notice that dust extinction will seriously affect the metallicity
calibration from the line ratio of [N~\textsc{ii}]/[O~\textsc{ii}] since the
blue line [O~\textsc{ii}] is strongly affected by dust extinction. 
Fig.~\ref{fig6} shows the results without extinction correction for 
[N~\textsc{ii}]/[O~\textsc{ii}] for our sample galaxies, which shows a much wider
distribution, and is quite far moved from KD02's model results. The
comparison between Fig.~\ref{fig6} and Fig.~\ref{fig4}c shows that: 1) the
dust extinction in these SDSS galaxies could not be omitted; 2) the
extinction coefficients of the individual galaxies are quite different,
thus, the data can show large scatter when the dust extinction inside the
individual galaxies were not estimated properly. For other ratios, like 
([O~\textsc{iii}]/H$\beta $)/([N~\textsc{ii}]/H$\alpha $), 
[N~\textsc{ii}]/H$\alpha $, [N~\textsc{ii}]/[S~\textsc{ii}] 
and[O~\textsc{iii}]/H$\beta $,
dust extinction does not have much effect on them due to the close positions
of the two related lines in wavelength. We did not consider the dust
extinction for [N~\textsc{ii}]/H$\alpha $ and [N~\textsc{ii}]/[S~\textsc{ii}
] ratios in our calculations.

\section{Nitrogen-to-oxygen abundance ratios}

It is possible to estimate the N abundances of galaxies from strong  optical
emission lines, which can help to understand the origin of nitrogen.

The basic nuclear mechanism to produce nitrogen is well understood -- it 
must result from CNO processing of oxygen and carbon in hydrogen burning. 
However, the nucleosynthetic origin of nitrogen has a ``primary" and a 
``secondary" component, which is still in debate. If the ``seed" oxygen  and
carbon are those incorporated into a star at its formation and a  constant
mass fraction is processed, then the amount of nitrogen produced  is
proportional to the initial heavy-element abundance, and the nitrogen 
synthesis is said to be ``secondary". If the oxygen and carbon are produced 
in the star prior to the CNO cycling (e.g. by helium burning in a core, 
followed by CNO cycling of this material mixed into a hydrogen-burning 
shell), then the amount of nitrogen produced may be fairly independent of 
the initial heavy-element abundance of the star, and the synthesis is said 
to be ``primary" (Vila-Costas \& Edmunds 1993). In general, primary 
nitrogen production is independent of metallicity, while secondary 
production is a linear function of it.

From a theoretical point of view, the $secondary$ production of nitrogen
should be common to stars of all masses, whereas the $primary$ production
should arise only from intermediate-mass stars ($4<M/M_{\odot }<8$)
undergoing dredge-up episodes during the asymptotic giant branch evolution.
In particular, when the third dredge-up is operating in conjunction with the
burning at the base of a convective envelope (hot-bottom burning), primary
nitrogen can originate (Renzini \& Voli 1981; Matteucci 1986). The large set
of database of galaxies from SDSS must provide important information on the
nucleosynthetic origin of nitrogen. The N/O ratio as a function of O/H is
the basic method to study the N abundance of galaxies.

\subsection{The method and results}

Firstly, we use the formula given by Thurston et al. (1996) to  estimate the
electron temperature in the [N~\textsc{ii}] emission region  
($T_{[\mathrm{NII]}}$) from log$R_{23}$:   
\begin{equation}
T_{[NII]}=6065+1600x+1878x^2+2803x^3,
\end{equation}
where $x$=log(R$_{23}$), $T_{[\mathrm{NII]}}$ is in units of K.
This formula was derived from the data points with about
-0.6$<$log$R_{23}$$<$0.8, and 5000K$<$$T_{[\mathrm{NII]}}$$<$10000K (see fig.1 of
Thurston et al. 1996). 

Then, log(N/O) values are  estimated from the 
([N~\textsc{ii}] $\lambda\lambda{6548},{6583}$)/([O~\textsc{ii}]$\lambda {3727}$) 
emission-line ratio and $T_{[\mathrm{NII]}}$ temperature by assuming 
$\frac{\mathrm{N}}{\mathrm{O}}=\frac{\mathrm{N^{+}}}{\mathrm{O^{+}}}$, 
and using the convenient
formula based  upon a five-level atom calculation given by Pagel et al.
(1992)  and Thurston et al. (1996):   
\begin{eqnarray}
&\mathrm{log ({\frac{{N^+}}{{O^+}}})}& 
= \mathrm{ log {\frac{{[NII]6548,6583}}{{[OII]3727}}}+0.307 -0.02log} t_{[NII]}  \nonumber \\
& & -\frac{{0.726}}{{t_{[NII]}}},
\end{eqnarray}
where $t_{[NII]}$ = 10$^{-4}$$T_{[\mathrm{NII]}}$.
We consider the flux of [N~\textsc{ii}]$\lambda{6548}$ is equal to 
one-third (0.333) of that of the [N~\textsc{ii}]$\lambda{6583}$ in calculations. 

\begin{figure}[tbp]
\centering
\includegraphics[width=7.8cm]{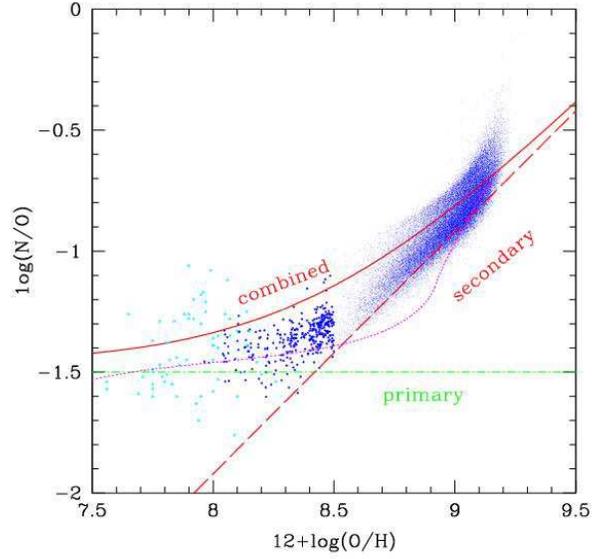}  
\caption{log(N/O) abundance ratios of our SDSS sample galaxies as functions
of their 12+log(O/H) abundances derived from $R_{23}$ by using the
calibration given by Tremonti et al. (2004). $The$ $small~ blue~ points$
refer to the observational data on metal-rich branch, and $the$ $stars$
refer to the SDSS sample galaxies with lower metallicities,
12+log(O/H)=8.0-8.5, which are the Bayesian metallicity estimates from
Tremonti et al. (2004) and are in the ``transition region" from the
metal-poor to the metal-rich branches. The $open~ triangles$ in cyan color
represent the 67 H~\textsc{ii} regions in 21 dwarf irregular galaxies taken
from van Zee \& Haynes (2006). The ``primary" (the dot-dashed line),
``secondary" (the long-dashed line) components and the combination of these
two components (the solid line) taken from Vila-Costas \& Edmunds (1993)
have also been plotted. The dotted line refers to the prediction of a
numerical one-zone model from Henry et al. (2000). }
\label{fig7}
\end{figure}

Fig.~\ref{fig7} gives the log(N/O) vs. 12+log(O/H) relations for our sample
galaxies. 
We also plot models for the primary and secondary origin of nitrogen 
taken from Vila-Costas \& Edmunds (1993):
the dot-dashed line refers to the
\textquotedblleft primary" component, 
the long-dashed line refers to
the \textquotedblleft secondary" component,
and the solid line refers to the combination of the two
components. The
log(N/O) abundances of the sample galaxies are consistent with the
combination of the \textquotedblleft primary" and \textquotedblleft
secondary" components, but the $secondary$ one dominates in these
metal-rich galaxies. This result is similar to the previous studies, e.g.
Shields et al. (1991), Vila-Costas \& Edmunds (1993), Contini et al. (2002),
and Kennicutt et al. (2003). In addition, 67 H~\textsc{ii} regions in 21
dwarf irregular galaxies taken from van Zee \& Haynes (2006) are also
plotted in Fig.~\ref{fig7} as triangles. The log(N/O)
abundances of these dwarf irregulars are dominated by the \textquotedblleft
primary" component. The SDSS galaxies with 8.0$<$12+log(O/H)$<$8.5 in the
metallicity transition region are also presented as the $stars$ in Fig.~\ref
{fig7}. They are just distributed in the region between the H~\textsc{ii}
regions in dwarf irregulars and the metal-rich galaxies.

\subsection{The analyses and discussions}

The SDSS sample galaxies show increasing log(N/O) abundance ratios following
increasing 12+log(O/H) abundances. The behavior of N/O with increasing
metallicity provides clues about the chemical evolution history of the
galaxies and the stellar populations responsible for producing oxygen and
nitrogen. Oxygen is mainly produced in short-lived, massive stars, and is
ejected to the ISM by Type II supernova explosions (SNe). In addition,
nitrogen is mainly produced in the long-lived, intermediate- and low-mass
stars, and is ejected to ISM through stellar wind. In particular, the
burning at the base of the convective envelope (hot-bottom burning) and in
conjunction with the third dredge-up process in intermediate-mass stars
contribute an important fraction of nitrogen (Renzini \& Voli 1981; van den
Hoek \& Groenewegen 1997; Liang et al. 2001; Henry et al. 2000). Therefore,
nitrogen will be delayed in its release into the ISM as compared to oxygen.

Contini et al. (2002) analyzed the delayed release of nitrogen. As they
discussed, during a long period of quiescence, intermediate-mass star
evolution will significantly enrich the galaxy in nitrogen but not in
oxygen; during starburst, however, N/O drops while O/H increases as the most
massive stars begin to die and supernovae release oxygen into the ISM. A few
tens of Myr after the burst, the massive stars producing oxygen will be
gone. Then N/O will rise again as intermediate-mass stars begin to
contribute to the primary and secondary production of nitrogen. A galaxy
undergoing successive starbursts will make the N/O ratio increase following
the increasing O/H abundance. However, the increasing N/O following O/H in
the metal-rich region does not have to require star formation to occur in
periodic starbursts. Rather, as long as the recent star formation activity
is less than the past star formation rate, the delayed release of nitrogen
from the aggregate intermediate mass stellar population will slowly increase
the N/O ratio since it is not balanced by additional production of oxygen in
high mass stars (van Zee \& Haynes 2006).

\begin{figure}[tbp]
\centering
\includegraphics[width=7.8cm]{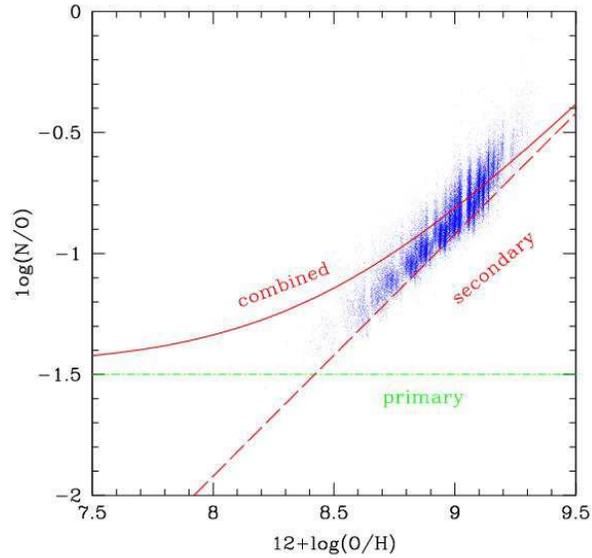}  
\caption{log(N/O) abundance ratios of the SDSS sample galaxies as a function
of the Bayesian metallicity estimates 12+log(O/H)$_{T04}$ from Tremonti et
al. (2004). The lines are the same as in Fig.~\protect\ref{fig7}. }
\label{fig8}
\end{figure}

One fact in Fig.~\ref{fig7} should be noticed: when the O/H increases to be
12+log(O/H)$\sim $9.0, the slope of the increase of log(N/O) becomes
steeper, i.e. the log(N/O) shows an upturn trend at higher metallicities
than 12+log(O/H)$\sim $9.0, or O/H then saturates for the increasing N/O.
The main reason for this may be that the sensitivity of $R_{23}$ to
metallicity saturates somewhat at a high metallicity. To understand more
about this, we rederive the relations of log(N/O) versus the other two sets
of $R_{23}$-derived oxygen abundances, 12+log(O/H)$_{K99}$ and 12+log(O/H)$
_{Z94}$. The later one is derived from the conversion of $R_{23}$ to
12+log(O/H) given by Zaritsky et al. (1994). They show the same trend as
12+log(O/H)$_{R_{23}}$. We then take the Bayesian estimates of 12+log(O/H)$_{T04}$
from Tremonti et al. (2004) to replace the $R_{23}$-derived oxygen
abundances to present the N/O vs. O/H relations of the sample galaxies. The
results are given in Fig.~\ref{fig8}. The saturation of 12+log(O/H) to
log(N/O) becomes much weaker now. This confirms that the saturation of O/H
to N/O at the metal-rich end mainly comes from the insensitivity of $R_{23}$
to 12+log(O/H) there. Indeed, our Fig.~\ref{fig1} has shown this saturation
of $R_{23}$-derived oxygen abundances at the high metallicity end.
That may mean that the upturn in N/O at high abundance is not physical.

From the comparisons above, we should note that care should be taken in
interpreting features in the N/O vs. O/H diagram because of systematic
uncertainties in metallicity calibration. If we use $T_{e}$ rather than
strong-line calibrations, O/H would be lower but N/O would not be much
affected since it has only a weak dependence on $T_{e}$. The net effect
would be a shift of the data to the left in Figs.~\ref{fig7},\ref{fig8}.
The data points will then be more consistent with 
the H~\textsc{ii} regions in M101 studied by Kennicutt et al. (2003) in the
relations of N/O vs. O/H diagram. The magenta dotted line in Fig.~\ref{fig7}
is the \textquotedblleft best" model B of the numerical one-zone models for
the evolution of N/O versus O/H given by Henry et al. (2000). Kennicutt et
al. (2003) found that the predicted N/O mode by this model is too low at
intermediate abundance (12+log(O/H)$_{T_{e}}$=8.2-8.6) to predict the
observed results of the H~\textsc{ii} regions in M101. Our SDSS sample
galaxies show a similar discrepancy from the model: at the intermediate
abundance 12+log(O/H)$<$9.0, the model predicts too low N/O at a
given O/H; for the 12+log(O/H)$>$9.0, although the model predicts an
increasing trend of N/O following O/H, the predicted N/O is still generally
lower than the observed results at a given O/H. Indeed, Henry et al. (2000)
themselves already pointed out that their model B provides a fit to the NO
envelope compared with the observational data (see their fig.3b).

In addition, dust extinction is a very important factor affecting the 
estimated (N/O) abundances, as well as the O/H abundances. It can
be understood from Fig.~\ref{fig6} (by comparing with Fig.~\ref{fig4}c). The dust
extinction of these SDSS sample galaxies has been considered when deriving
their abundances. Moreover, it is a homogeneous sample taken from one single
facility using the same set-up, which minimizes the systematic errors from
observation runs, and the [N~\textsc{ii}]/[O~\textsc{ii}] ratio is almost
independent of the ionization parameter (KD02). Therefore, these SDSS data
points would not show much spread in the N/O vs O/H relations due to
observational errors. The
observational scatter of these galaxies are about 0.062\thinspace dex
estimated from Fig.~\ref{fig8}.

\begin{figure}[tbp]
\centering
\includegraphics[width=7.8cm]{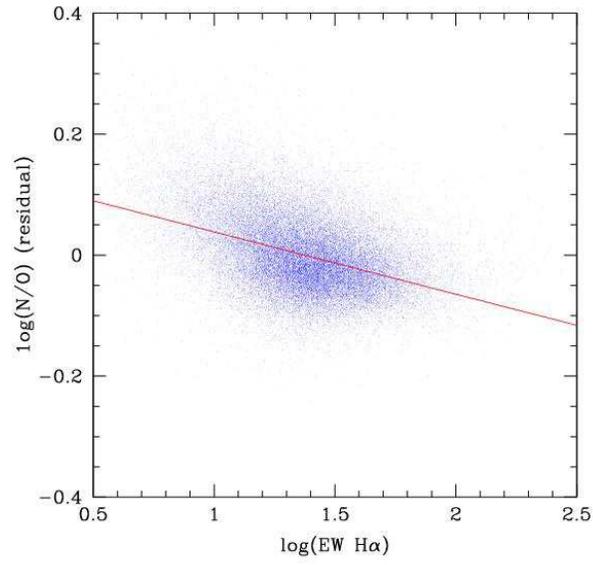}   
\caption{log(N/O) abundance residuals of our SDSS sample galaxies as
a function of the equivalent widths of their H$\protect\alpha$ emission lines
EWs(H$\protect\alpha$) (in \AA ). The solid line refers to a linear 
least-squares fit with a rms $\sim$0.057 for the data points (Eq.(5)). }
\label{fig9}
\end{figure}

 The scatter in the observed log(N/O) vs. log(O/H) relation is
 relatively small. Nevertheless, we can use our large
 dataset to see if the residuals correlate with any physical properties
 of galaxies. If the release of nitrogen is indeed delayed with
 respect to oxygen, we might expect to find a correlation between the
 N/O residuals and the stellar birthrate, which is the ratio of the
 present to past average star formation rate (SFR).
 The equivalent widths of H$\alpha $ emission lines, log(EW(H$\alpha $)),
 provides a good proxy
 for the stellar birthrate (Kennicutt et al. 1994).
 Fig.~\ref{fig9}
shows the relation between log(N/O)$_{\rm residual}$ and the 
log(EW(H$\alpha $)) for the sample galaxies. 
log(N/O)$_{\rm residual}$ is the N/O abundance
having its dependence on O/H abundances been removed. To do so, first we obtain a
linear least squares fit for the relation between log(N/O) vs. 
12+log(O/H)$_{T04}$ for the sample galaxies, then the observed log(N/O) 
abundances were subtracted by the fitted values predicted by this 
linear relation. Fig.~\ref{fig9} shows that the log(N/O) abundances decrease 
following the increasing EWs(H$\alpha $) for these galaxies. 
The solid line refers to a linear
least squares fit for this observed relation with a rms $\sim $0.057: 
\begin{equation}
\mathrm{log({\frac{{N}}{{O}}})_{residual}=-0.102\times log(EW(H\alpha ))+0.140.}
\end{equation}
This relation means that the log(N/O) increases following the decreasing
star formation rates, which is consistent with the suggestion of van Zee \&
Haynes (2006) for the relation between log(N/O) and B-V colors for their
dwarf galaxies. They suggested that the high N/O ratio are correlated with
redder systems and decreasing star formation rates.

\section{Conclusion}

We have use the observational data of a large sample of 38,478 star-forming
galaxies selected from SDSS-DR2 to derive oxygen abundance calibrations of
12+log(O/H) versus several metallicity-sensitive emission-line ratios,
including [N~\textsc{ii}]/H$\alpha $, [O~\textsc{iii}]/[N~\textsc{ii}], 
[N~\textsc{ii}]/[O~\textsc{ii}], [N~\textsc{ii}]/[S~\textsc{ii}], 
[S~\textsc{ii}]/H$\alpha $, and [O~\textsc{iii}]/H$\beta $. 
These calibrations are very
useful when the \textquotedblleft $T_{e}$-method", the most accurate method,
and the empirical \textquotedblleft $R_{23}$-method" are not available to
estimate oxygen abundances of galaxies. In addition, they can overcome the
\textquotedblleft double-valued" drawback of the  $R_{23}$-method.  

We fit the observed relations with a linear least squares fit
and/or a third-order polynomial fit. All the fitting coefficients and the
rms derivations (0.032-0.078\thinspace dex) have been given in
Table\thinspace 1. 
When we only select the sample galaxies with higher
signal-to-noise ratio emission lines, e.g. $>$10$\sigma $, the rms
derivations only have small changes, which means that probably most of the
scatter is intrinsic and not an artifact of observational errors. 
The scattering of the data points mainly come from the
different ionization parameters in the individual galaxies, which is
confirmed by the smallest rms value of [N~\textsc{ii}]/[O~\textsc{ii}]
ratio, which is independent on the ionization parameters. Such a large
sample of galaxies can show some information never before obtained from much
smaller samples, for example, the turnover of the relations between
12+log(O/H) versus [N~\textsc{ii}]/H$\alpha $ at 12+log(O/H)$>$9.0. 

 Among these calibrations, the [N~\textsc{ii}]/[O~\textsc{ii}] ratio
should be the best one for metallicity calibration in the metal-rich branch
(same as KD02) because it shows a monotonical increase following the
increasing metallicity and less scatter than other line ratios, and it is
independent of the ionization parameter. However, dust extinction must be
estimated properly before using this calibration indicator because of the
blue line [O~\textsc{ii}] and the far positions of the two lines in
wavelength. ([O~\textsc{iii}]/H$\beta $)/([N~\textsc{ii}]/H$\alpha $) is
also a good indicator, a bit better than [N~\textsc{ii}]/[S~\textsc{ii}] and
[O~\textsc{iii}]/H$\beta $ due to the relatively smaller rms derivation. The
discrepancy between the data and models for [N~\textsc{ii}]/[S~\textsc{ii}]
highlights the need for empirical calibrations. [N~\textsc{ii}]/H$\alpha $
is only a good indicator for the galaxies with 12+log(O/H)$<$9.0, and the
problem at higher metallicities is the turnover of [N~\textsc{ii}]/H$\alpha $
following increasing O/H. However, [N~\textsc{ii}]/H$\alpha $ may be one of
the most useful calibrations since it is not affected by dust extinction due
to the very close positions of the two related lines in wavelength, and can
be detected by the near infrared instruments for the intermediate- and high-$
z$ star-forming galaxies. We should notice that the derived calibration will
result in a higher 12+log(O/H) abundance at a given [N~\textsc{ii}]/H$\alpha 
$ ratio than those of Pettini \& Pagel (2004) and Denicol\'{o} et al.
(2002). The main reason for this difference may be due to the discrepancy
between the metallicity estimates from the $T_{e}$-method and 
$R_{23}$-method.

 The resulting calibrations are more suitable for luminous and
metal-rich galaxies since they have metallicities in the range 8.4$<$
12+log(O/H)$_{R_{23}}$ $<$9.3.
Because our calibrations are derived from SDSS data, they will be
best suited for data which samples the inner few kiloparsecs of
 galaxies rather than the whole disk. This aperture effect is most
 significant for line ratios with a strong dependence on ionization
 parameter (e.g. [N~\textsc{ii}]/[O~\textsc{iii}]).

The observed calibration relations are compared with the photoionization
models from KD02, which generally show good consistent trends, including the
turnover trend of [N~\textsc{ii}]/H$\alpha $ at the metal-rich end, the
independence of [N~\textsc{ii}]/[O~\textsc{ii}] on ionization parameter,
etc. However, most of these sample galaxies are distributed in a narrower
range of the ionization parameter $q$ than the models. The actual range of 
$q$ is from $1\times 10^{7}$ to $8\times 10^{7}$ cm\thinspace s$^{-1}$, i.e.,
log$U$=-3.5 to -2.5, but the models cover from $q=5\times 10^{6}$ to 
$3\times 10^{8}$ cm\thinspace s$^{-1}$, i.e., log$U$=-3.78 to -2. 

Another contribution of this work is that we obtained the log(N/O) abundance
ratios for this large sample of SDSS star-forming galaxies with 8.4$<$
12+log(O/H)$<$9.3. Their log(N/O) abundances are consistent with the
combination of the \textquotedblleft primary" and \textquotedblleft
secondary" components of nitrogen, but the \textquotedblleft secondary" one
dominates at high metallicities. The increasing log(N/O) abundances
that follow the increasing metallicities for these metal-rich galaxies can
be the prediction of continuous but declining star formation rates, which is
confirmed by the linear relation with a negative slope of the 
log(N/O)$_{\rm residual}$ versus log(EW(H$\alpha $)). 
The scatter of these observational data
are $\sim $0.062\thinspace dex in the log(N/O) versus 12+log(O/H) relations.
However, we should notice that dust extinction strongly affects the
estimates of the log(N/O) abundance ratios and the corresponding scatter for
the sample galaxies.

In summary, this work provides several useful metallicity calibrations based
on strong emission-line ratios of 38,478 SDSS star-forming galaxies, and
estimates the log(N/O) abundance ratios for such a large number of galaxies.
These can be used as references for future studies of metallicities of
galaxies. We should notice that, 
when these calibrations are used for high redshift objects,
the indices based on [N~\textsc{ii}] may have some extra uncertainties
since for very high redshifts N/O vs. O/H may be different from the local case
because the galaxies are not old enough to produce significant 
amounts of secondary nitrogen.

\acknowledgments
We specially appreciate the referee for the many valuable and wise
comments and suggestions, which greatly help us in improving this
work.  We thank Rob Kennicutt for his wise suggestions in science
and the English edition which have improved the paper well.
We thank James Wicker and  Richard de Grijs for their warm help
in improving the English description, and Chang  Ruixiang, Shen Shiying, Liu
Chengze, Wu Hong, Xu Yan, and Cao Chen for the helpful discussions
about the SDSS database. We thank Lisa Kewley for the private
communication that has helped us to understand better their models. 
We thank Max Pettini for his important comments on this work, and 
Richard C. Henry for sending us his chemical evolution model results
for N/O  vs. O/H. This work was supported by the Natural Science
Foundation of China  (NSFC) Foundation under No.10403006, 10433010
10373005, 10573022, 10333060, and 10521001.




 \clearpage

 \clearpage

 \clearpage

\clearpage
\clearpage

\begin{table}[h!]
\caption{{\protect\scriptsize Coefficients for the oxygen abundance
diagnostics. Defining oh=12+log(O/H) here; assuming y=b0+b1x, and
y=a0+a1x+a2x$^2$+a3x$^3$ for the linear and 3-order polynomial fits,
respectively. Here, y=log$R_{23}$, 12+log(O/H)$_{R_{23}}$, 12+log(O/H)$_{T04}
$ and 12+log(O/H)$_{K99}$ for the different cases, and x= N2, O3N2, N2O2,
N2S2, O3Hb for each of the ``y" case. $\protect\sigma$ is the rms deviation
from the data to the fitted line in dex.}}
\label{tab1}
\begin{center}
{\scriptsize \ 
\begin{tabular}{lcccccc}
\hline\hline
(1) & (2) & (3) & (4) & (5) & (6) & (7) \\ \hline
diagnostics & N2 & N2 & O3N2 & N2O2 & N2S2 & O3Hb \\ \cline{2-7}
&  &  &  &  &  &  \\ 
& log${\frac{{[NII]6583}}{{H\alpha}}}$ & log${\frac{{[NII]6583}}{{H\alpha}}}$
& log${\frac{{[OIII]5007/H\beta}}{{[NII]6583/H\alpha}}}$ & log${\frac{{
[NII]6583}}{{[OII]3727}}}$ & log${\frac{{[NII]6583}}{{[SII]6717,6731}}}$ & 
log${\frac{{[OIII]4959,5007}}{{H\beta}}}$ \\ \cline{2-7}
appropriate range & [-1.2,-0.55] & [-1.2,-0.2] & [-0.7,1.6] & [-1.4,0.7] & 
[-0.5,0.8] & [-0.9,0.6] \\ \cline{3-7}
& 8.4$<$oh$<$9.0 & \multicolumn{5}{c}{8.4$<$oh$<$9.3} \\ \hline
 &  &  & Linear  fits & &  &  \\ 
\textbf{12+log(O/H)$_{R_{23}}$} &  &  &  &  &  &  \\ 
b0 & ... & ... & 9.010 & 9.125 & 8.944 & 8.846 \\ 
b1 & ... & ... & -0.373 & 0.490 & 0.669 & -0.497 \\ 
$\sigma$(dex) & ... & ... & 0.059 & 0.039 & 0.075 & 0.062 \\  \hline
\textbf{12+log(O/H)$_{T04}$} &  &  &  &  &  &  \\ 
b0 & ... & ... & 9.007 & 9.128 & 8.934 & 8.845 \\ 
b1 & ... & ... & -0.384 & 0.514 & 0.789 & -0.490 \\ 
$\sigma$(dex) & ... & ... & 0.067 & 0.041 & 0.050 & 0.078 \\  \hline
\textbf{12+log(O/H)$_{K99}$} &  &  &  &  &  &  \\ 
b0 & ... & ... & 8.916 & 9.020 & 8.857 & 8.771 \\ 
b1 & ... & ... & -0.332 & 0.443 & 0.605 & -0.441 \\ 
$\sigma$(dex) & ... & ... & 0.058 & 0.037 & 0.068 & 0.061 \\  \hline \hline
 &  &  & Polynomial fits &  &  &  \\ 
\textbf{12+log(O/H)$_{R_{23}}$} &  &  &  &  &  &  \\ 
a0 & 10.305 & 11.948 & 9.013 & 9.118 & 8.956 & 8.841 \\ 
a1 & 4.786 & 11.843 & -0.395 & 0.338 & 0.746 & -0.575 \\ 
a2 & 5.079 & 14.986 & -0.072 & -0.319 & -0.472 & 0.011 \\ 
a3 & 2.191 & 6.747 & 0.106 & -0.125 & -0.807 & 0.297 \\ 
$\sigma$(dex) & 0.063 & 0.058 & 0.058 & 0.035 & 0.072 & 0.061 \\ \hline
\textbf{12+log(O/H)$_{T04}$} &  &  &  &  &  &  \\ 
a0 & 9.617 & 11.835 & 9.011 & 9.121 & 8.945 & 8.839 \\ 
a1 & 1.563 & 11.317 & -0.406 & 0.384 & 0.838 & -0.584 \\ 
a2 & 0.262 & 14.289 & -0.086 & -0.229 & -0.380 & 0.001 \\ 
a3 & -0.048 & 6.566 & 0.119 & -0.052 & -0.481 & 0.342 \\ 
$\sigma$(dex) & 0.051 & 0.050 & 0.066 & 0.038 & 0.048 & 0.076 \\ \hline
\textbf{12+log(O/H)$_{K99}$} &  &  &  &  &  &  \\ 
a0 & 12.074 & 12.464 & 8.919 & 9.014 & 8.868 & 8.762 \\ 
a1 & 14.410 & 15.612 & -0.358 & 0.287 & 0.682 & -0.514 \\ 
a2 & 21.398 & 22.305 & -0.085 & -0.362 & -0.423 & 0.076 \\ 
a3 & 11.193 & 11.185 & 0.126 & -0.171 & -0.842 & 0.368 \\ 
$\sigma$(dex) & 0.070 & 0.065 & 0.057 & 0.032 & 0.066 & 0.059 \\ \hline
\end{tabular}
}
\end{center}
\par
\end{table}


\end{document}